\documentclass[12pt]{article}
\usepackage{graphicx}
\usepackage[top=3cm, bottom=2cm, left=3cm, right=3cm]{geometry}

\title{The Diurnal Temperature Range for Europe - \\a search for Cosmic Ray Forbush 
Decrease manifestations and the DTR periodicities.}
\author{A. D. Erlykin $^{1,2}$ and A. W. Wolfendale $^{2}$ \\
$(1)$ P.N.Lebedev Physical Institute, Moscow, Russia \\
corresponding author: tel.+74991358737, e-mail: erlykin@sci.lebedev.ru\\
$(2)$ Department of Physics, Durham University, Durham, UK}

\begin{document}
\maketitle

\begin{abstract}
Following on previous work by others, which gave evidence for few-day changes in the 
European Diurnal Temperature Range (DTR) apparently correlated with Cosmic Ray Forbush 
Decreases, we have made an independent study. We find no positive evidence.

An analysis has also been made of the Fourier components of the time series of the DTR 
value (taken as deviations from a $\pm$10 day running mean).
Evidence for a number of interesting periods is found, including one at about 27 days, 
albeit with a variability with time. The same period of solar irradiance (particularly 
in the UV) is favoured as the explanation.
\end{abstract}
\section{Introduction}
Uncertainty relating to the relevance of Cosmic Rays (CR) to Climate Change continues. 
Although some work (our own included - eg Erlykin et al. \cite{EW1,EW2,EW3} has cast 
doubts on the claims of Svensmark and Friis-Christensen \cite{Sven1} and others that 
there is a causal correlation of Low Cloud Cover (LCC, altitude $<$3.2 km, ie the 
`boundary layer') with CR intensity, doubts remain. The supporters draw attention to 
the results from the CLOUD project \cite{Kirk} which give evidence for an effect of 
accelerator-particles on nucleation rates.

The supporters also draw attention to the Diurnal Temperature Range (DTR) data from 
European stations where there is an apparent correlation with Forbush Decreases (FD) 
according to  Dragic et al. \cite{Drag}.

The work just referred to is in the spirit of studies by Lu and Sanche \cite{LuSa}, and
 others, which claimed evidence for CR having an effect on Atmospheric 
Chlorofluorcarbon Dissociation and Ozone depletion in the Polar atmosphere. It is with 
the DTR work that we are concerned here.

Although the DTR, which is the day-night temperature difference, is affected by a 
number of atmospheric parameters (cloud cover, precipitation, changes in agriculture, 
water vapour feedback etc) it is an important quantity to study for two reasons.
\begin{itemize}
\item[(i)]It can be measured in a straightforward way, and
\item[(ii)]It decreased on a global scale from the 1950s onwards, at the time when the 
mean Global temperature started to increase markedly (`Global Warming'). DTR has in it 
therefore a number of (often dependent) climate parameters and these could be, in 
principle, CR-dependent (see \cite{Mak} for details of DTR for Europe for the range 
1950-2005).
\end{itemize}

FD analyses are useful in that for the short time spans involved (days) phenomena other
 than CR might be expected to change only slowly so that CR effects dominate.
\section{Previous FD studies}
Searches for the correlation of FD with other atmospheric parameters have already been 
made. Specifically, Svensmark et al. \cite{Sven2} searched for, and found, changes 
correlated
 with strong FD for the atmospheric liquid water fraction (LCF) - as detected by 
`MODIS', and Low IR cloud cover over regions - as recorded by the `ISCCP', together 
with aerosol content (using AERONET) and  the cloud water content using SSM/1. The 
results were all similar: the FD was followed within a few days by similar changes in 
the atmospheric properties.

In the event, Laken et al. \cite{Lak1}, concentrating on the LCF (MODIS), found no 
support for the hypothesis. However, the topic is so important that we have repeated 
the analysis, but this time using the DTR as the atmospheric parameter under 
consideration.

This is not the first time such a study has been made. As mentioned already, Dragic et 
al. \cite{Drag} found a significant correlation for a limited class of FD: those above 
7\% in magnitude. 
The DTR data were taken from 189 European meteorological stations; the stations were 
randomly selected but covered the entire European region. Although details will be 
given later, the `result' can be mentioned here: an increase in mean DTR for FD$>$7\% 
of (0.38$\pm$0.06)$^{\circ}$C. Taken at its face value the result is very significant.

Although the results used are only for European stations, and are thus not Global in 
any sense, they may, nevertheless, be very useful. As Lockwood \cite{Lock} remarks in 
his comprehensive review of solar influences on Global and Regional climates, the 
latter are much more significant than the former.
\section{The present analysis of FD versus DTR.}
\subsection{Sources of the data.}
The data, comprising the daily sequence of mean DTR and CR intensity, were provided by 
Dragic at al. \cite{Drag} and Laken (2012, private communication ).
\subsection{Processing of the data.}
There seems to be no contention as to their validity and their manipulation in that our
 own work confirmed that of Dragic at al. 
\cite{Drag}, at least in a general way. 
Figure 1 gives the result from our own analysis for FD $>$7\%. It should be mentioned 
that the amplitude of the effect presented in Figure 1 is smaller and the uncertainties
 - errors - presented in Figure 1 are bigger than those given by Dragic et al. 
\cite{Drag}, viz. 0.25$\pm$0.10$^\circ$C, to be compared with 0.38$\pm$0.06$^\circ$C 
(~see \S2~); this topic is taken up again later. The argument is the extent to which 
the profile can be regarded as significant.
\begin{figure}[hptb]
\begin{center}
\includegraphics[width=10cm,height=15cm,angle=-90]{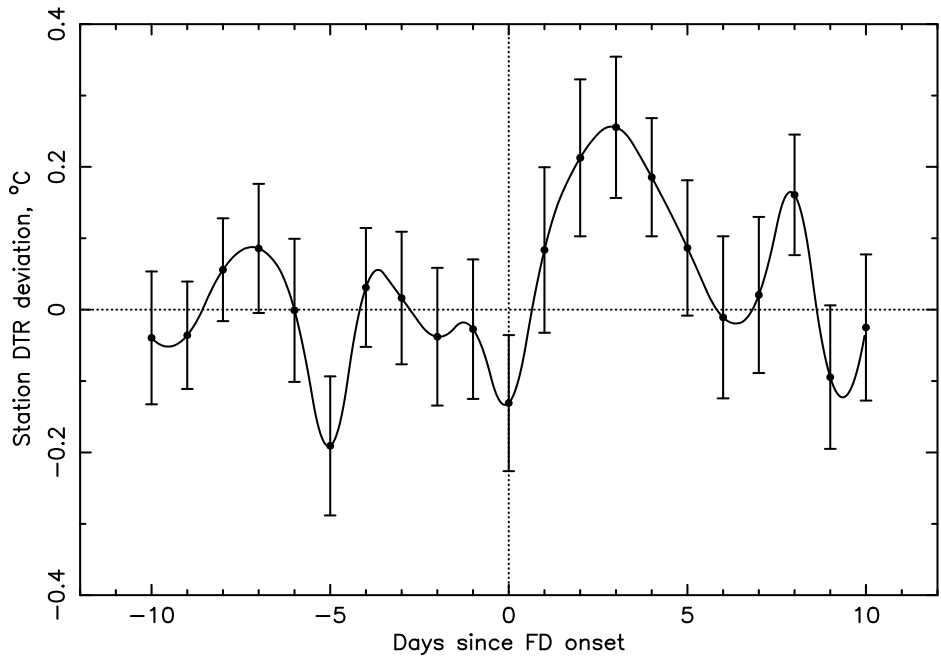}
\caption{\footnotesize The results showing the DTR signal following FD decreases. 
FD greater than 7\% were selected, of which there were 37. Dragic et al., \cite{Drag}, 
found a similar, though larger profile.}
\label{fig:fig1a}
\end{center}
\end{figure}

\subsection{Check on the chance of significance.}

\begin{figure}[ht]
\begin{center}
\includegraphics[width=10cm,height=15cm,angle=-90]{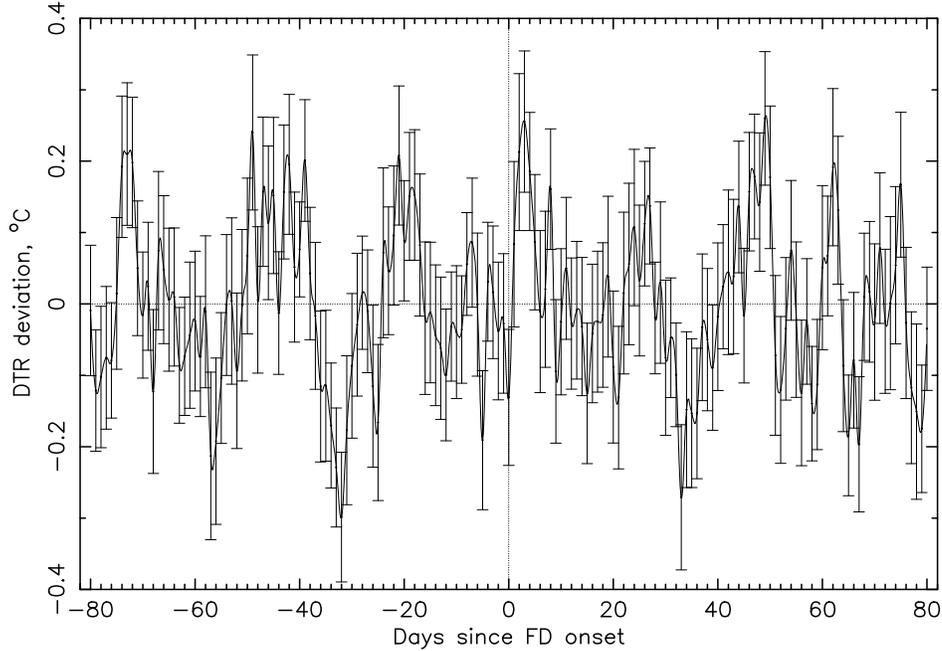}
\caption{\footnotesize Profile of mean DTR deviations from the 21-day running mean 
averaged over $\pm$80 days around the onset of 37 FD with magnitude $>$7\%}
\label{fig:fig2}
\end{center}
\end{figure}

Figure 2 gives the result of our analysis for the 37 events with FD $>$7\%; this 
relates to a wider time range than covered by Figure 1 (ie $\pm$80 days, to be compared
 with $\pm$10 days). For $\pm$10 days the plots coincide precisely with the plot given 
in Figure 1. It is immediately apparent that peaks similar to that for $\pm$10 days are
 not infrequent and, furthermore, the difference between pre- and post-FD is not 
significant. The problem is to determine the frequency of a peak which would be counted
 as acceptable, ie would be evidence for a (strong) FD causing a significant increase 
in the DTR. There are two parameters to be considered, as follows.
\begin{itemize}
\item[1.]The probability of a peak of acceptable height, and time width.
\item[2.]The phase lag between the FD onset and the onset of the DTR increase. In 
\cite{Drag} it was zero days for FD $>$7\%, +1 days for FD: 7-10\% and -1 days for FD $>$10\%. In \cite{Sven2} it was approximately +6 days.
\end{itemize}

Starting with (1) we have smoothed the DTR data with a 21-day running mean and studied 
the patterns following datum times where the deviation from the datum level, 
$\Delta$(DTR) was less than 0.05$^\circ$C.  We consider that 2 to 4 consecutive 
displacements above 0.2$^\circ$C would give patterns as acceptable evidence for a 
significant DTR signal (see Figures 1 \& 2). The results are given in Table 1. 
\begin{center}
\begin{tabular}{| c | c || c | c |}
    \hline
    $\Delta$2,  0.2$^{\circ}$C & 20\% & $\Delta$4,  0.2$^{\circ}$C & 8\%  \\
    $\Delta$2,  0.1$^{\circ}$C & 26\% & $\Delta$4,  0.1$^{\circ}$C & 12\% \\ \hline
    $\Delta$2, -0.2$^{\circ}$C & 22\% & $\Delta$4, -0.2$^{\circ}$C & 9\%  \\
    $\Delta$2, -0.1$^{\circ}$C & 29\% & $\Delta$4, -0.1$^{\circ}$C & 10\% \\
                           & ($\pm$2\%) &         & ($\pm$1\%)            \\ \hline
\end{tabular}
\end{center}
{\footnotesize Table 1. Analysis of the chance probability of 2 (or 4) 
consecutive values of the DTR, $\Delta$, being greater than 0.2$^{\circ}$C (or 
0.1$^{\circ}$C), 
denoted $\Delta$2, 0.2$^\circ$C; $\Delta$2, 0.1$^\circ$C, $\Delta$4, (0.2$^\circ$C), 
$\Delta$4, (0.1$^\circ$C). Values for negative excursions are also indicated, they are 
essentially the same as for the positive excursions.}

\vspace{2mm}

It is seen that a probability of about 15\% is indicated for zero time delay (phase 
lag).

Concerning the phase lag that would have been regarded as allowable, this appears to be
 at least 3 days. Thus, the overall probability of a chance pattern being accepted as 
genuine is about 45\%.

Moving on from Forbush Decreases we examine the periodicity in the DTR deviations; 
Figure 2 leads us to believe that there are one or more such periodicities.
\section{The periodicity of DTR excursions.}
\subsection{The analysis.}
The presence of a periodicity (~or periodicities~) could have relevence to solar 
effects, or 
related to CR effects. Periodicities in the region of 27 days might be expected in that
 this is the (mean) solar rotation period although it is appreciated that the rotation 
period varies from about 25 days at the Equator to 38 days at the Poles and a clear 
result may not be expected. 

In the whole data set
some 1.37x10$^4$ continuous daily values of the DTR deviations (from the 21 day running
 mean) are available. A Fourier analysis yielded the results shown in Figure 3. 
\begin{figure}[htb]
\begin{center}
\includegraphics[width=10cm,height=15cm,angle=-90]{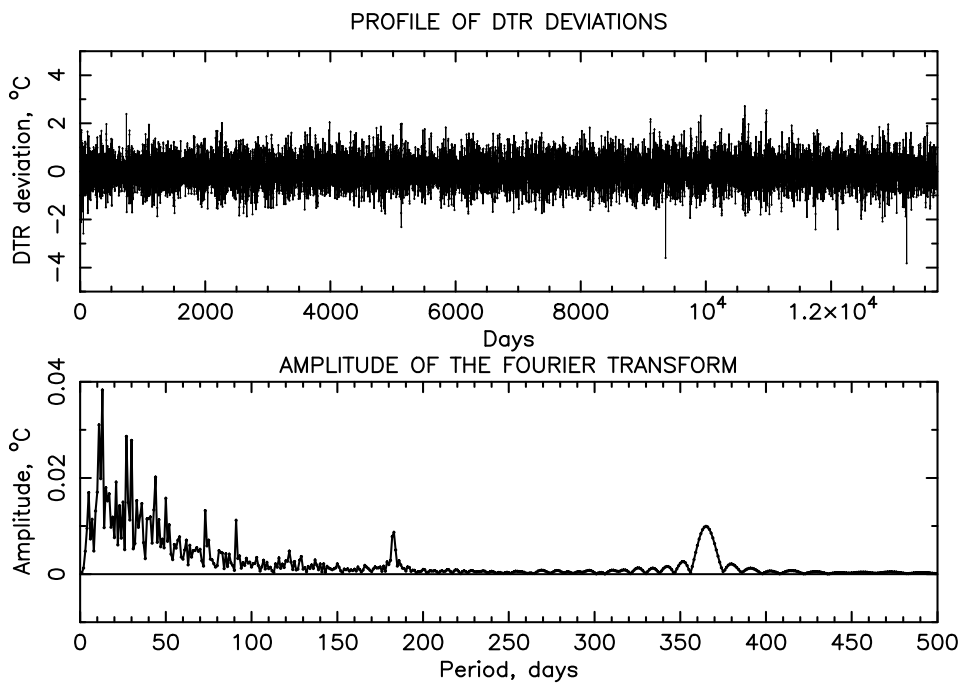}
\caption{\footnotesize Profile of DTR deviations from the 21-day running mean 
(~as for Figure 2~) for the whole period of 13,673 days, together with the amplitude of
 the Fourier transforms.}
\label{fig:fig3}
\end{center}
\end{figure}

The expected peaks at 12 months and 6 months are clearly visible, thus confirming the 
quality of the data. Other significant sharp peaks are present at the harmonics of the 
annual peak, viz one year divided by 3,4 etc. Physically interesting peaks are also 
present near 27 days and 13 days. In Table 2 we give the periods for the two most 
significant peaks in the profiles, for various data sets, as follows
\begin{itemize}
\item[(i)] 80 days before the FD events and 80 days afterwards (~Figure 2; the 
amplitudes of the Fourier transforms are not given~)
\item[(ii)] The whole period of observation (~Figure 3~)
\item[(iii)] Each solar cycle (~21, 22 and 23~) separately.
\end{itemize}
\begin{center}
\begin{tabular}{| c | l | l |}
    \hline
    Time range         &          &           \\ \hline
    80 days before FD  & 25(B)    &  11       \\ 
    80 days after FD   & 22(B)    &  12       \\
    (graphs not shown) &          &           \\ \hline
    13, 673 days       & 28,30    &  11,13(B) \\
    (the whole period) &          &           \\ \hline
    Cycle 21           & 22(B) 27 &           \\ \hline
    Cycle 22           & 30       & 13(B)     \\ \hline
    Cycle 23           & 28(B) 27 &           \\ \hline  
\end{tabular}
\end{center}
{\footnotesize Table 2. The two most significant peaks in the Fourier plots for 
different time ranges. 'B' denotes the bigger of the two. For close pairs both are 
given.}

\vspace{2mm}

It is evident from Table 2 that there is some grouping of the periods round 27d 
(~actually 22d to 30d~) and 13d (~actually 11d and 13d~).  To say that we have 
confirmed these periods would be too strong a statement but there is some evidence 
favouring them. In what follows an examination is given accepting the evidence as real.
 
\subsection{Interpretation of the possible 27d and 13d peaks}
A 27-day peak is interesting in that it is clearly solar-related, 27-day being the 
mean period of solar rotation, as remarked in \S4.1 It is well known that there is a 
near-27 day CR intensity variation (eg Dorman \cite{Dorm}) for which the amplitude is 
about 0.4\% \cite{Gil}. There is also a difference between alternate solar cycles. 
These features are attributed to energetic processes on the surface of the sun which 
give rise to time-dependent solar winds, and thereby periodicities of about 27 and 13 
days, the latter arising from frequent 180$^{\circ}$ longitude differences in the solar
 surface disturbances \cite{Pap}.

There is a problem with the solar-related periodicity analysis, however, due to the 
well-known fact that the solar rotation period is solar-latitude related. Thus, our
analysis of different lengths of time may give different periods. 
This topic has been followed up by examining the results cycle by cycle, viz for solar 
cycles 21, 22 and 23.

This is the reason for giving the division shown in Figure 2 for which details are 
given in Table 2. The results are rather disappointing in that there is no equality in 
the patterns for the alternate cycles: 21 and 23. Nevertheless, we maintain that there 
is some modest evidence favouring identification of solar-rotation features in the DTR 
record.
\begin{figure}[hptb]
\begin{center}
\includegraphics[width=10cm,height=15cm,angle=-90]{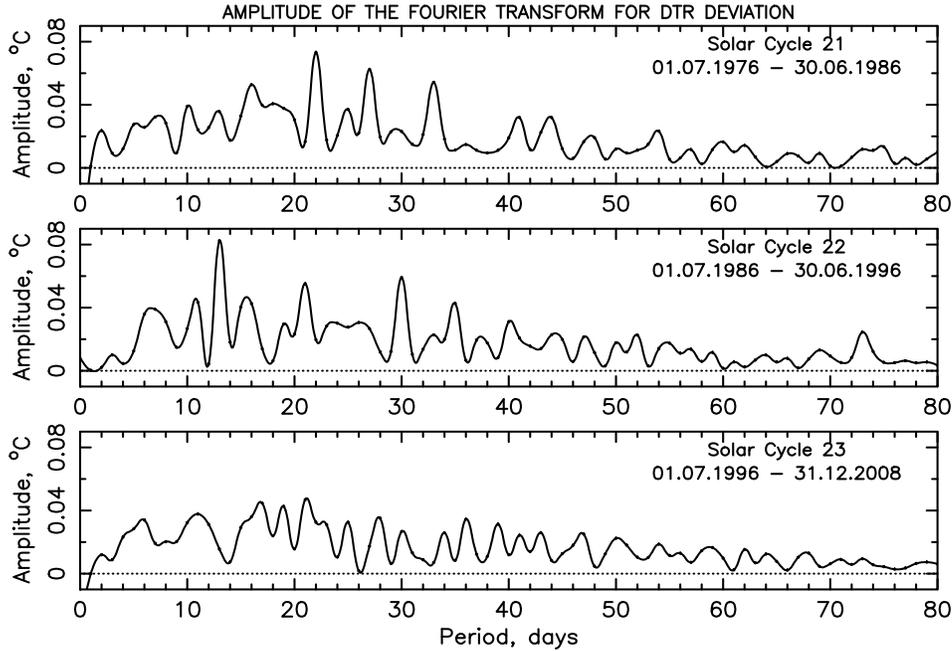}
\caption{\footnotesize Profiles of Fourier amplitudes for the Solar Cycles 21, 23 and 
23.}
\label{fig:fig4}
\end{center}
\end{figure}
\subsection{The quoted errors in the DTR measurements}
A good feature of Figure 2, which was used for periodicity studies, is that it allows a
 check on the veracity of our quoted errors. 
The number of (daily) values greater than 0, 1 and 2 standard deviations are: 360, 106 
and 11. For a Gaussian we expect 360, 120 and 18. It follows that, due to the data 
being deviations from a 
running mean, our 'errors' are slightly overestimated. Using the numbers just given the
 overestimate is calculated to be by 10\%. This is much smaller than the 70\% by which 
our errors exceeded those of Dragic et al. \cite{Drag} in their important paper which 
stimulated the present work. The origin of the difference is not known.
\section{The DTR-Cosmic Ray correlation - is it causal ?}
There is the standard problem of whether in the DTR the possible 13 and 27 day periods 
are Cosmic Ray related (~ie there is a causal connection~)., or whether both variations
 are caused by a third phenomenon. The alternatives here are therefore that the DTR 
periodicity is caused by
\begin{itemize}
\item[(i)]solar luminosity per se, which could cause climate change, or,
\item[(ii)]cosmic rays, via ionization affecting the degree of cloud cover.
\end{itemize}

It appears that no analysis of the results, alone, can make a distinction, because the 
Total Solar Irradiation (TSI) and Sunspot number are 
tightly correlated for obvious reasons and the Sunspot number and CR intensity are 
correlated by way of the variation in the Solar Wind.

Instead, recourse must be made to other results.

We consider that solar irradiance variations are a more likely source of the DTR 
periods than CR for the following reasons.
\begin{itemize}
\item[(i)]The source of the DTR change itself is claimed to be `changes in 
(terrestrial) emissions and the associated changes in incoming solar radiation' 
\cite{Mak}. Thus, changes in the solar irradiance are preferred. In this 
context, Rotman \cite{Rotm} has drawn attention to the very large amplitude of the 27 
day variation of the UV (120-300 nm) irradiance.
\item[(ii)]The CR ionization mechanism runs into serious difficulties. Although it is 
true that for Cycle 22 the CR Low Cloud Cover (LCC) correlation was strong \cite{Sven1}
  that for the next cycle was not (see, for example, \cite{EW1}. Most importantly, the 
last-mentioned authors drew attention to the fact
 that the CR and MCC are anti-correlated to the extent that the sum of LCC and MCC (and
 indeed the total CC) is not correlated with the CR. Thus, the role of CC in causing 
the ground-level irradiance variation which contributes to the DTR, CR correlation (at 
least for an 11-year period) is negligible.
\end{itemize}

Other serious problems for CR contributing via ionization to Cloud Cover come from the 
following:
\begin{itemize}
\item[(i)] Ionization caused by nuclear explosions, terrestrial radon and the Chernobyl
 nuclear accident had no effect on Cloud Cover \cite{EW2}.
\item[(ii)]Those clouds (stratiform), which should have a greater sensitivity to CR, do
 not show it \cite{EW3}.
\item[(iii)]Recent work at CLOUD \cite{Kirk} has shown that ionizing 
particles do produce nucleation under carefully controlled conditions, at least using 
the atmospherically relevant H$_2$SO$_4$. However, there is a very great temperature 
dependence of the rate and for most of the troposphere, and particularly the `boundary 
layer' - at altitudes below several km - the atmospheric temperatures are too high. 
Kirkby et al. agree with this diagnosis \cite{Kirk}.
\end{itemize}
The causal nature of the tropospheric CR, CC correlation on the 11-year scale and the 
DTR correlation on the four day (Forbush Decrease), 27 day and 1 year scales is thus 
strongly disfavoured. An explanation in terms of changes to the solar irradiance is 
more likely.

The dismissal of a causal connection for tropospheric climate parameters and CR does 
not mean that CR have no effect on the terrestrial atmosphere as a whole. There is 
evidence for CR effects in the stratosphere as mentioned in \S1. These are very small, 
however, and unlikely to be of much relevance to the lower atmosphere, changes in which
 are of such contemporary concern.

Finally, it should be noted that the slow increase in the CR intensity over the past 
two decades, as evinced by neutron monitors (eg that as on the Oulu web site 
\cite{Oulu}), shows that even if there were a CR, tropospheric atmosphere link it would
 not contribute to the observed current Climate Change.

\section{Conclusions.}
Concerning the effect of Forbush Decreases on the Diurnal Temperature Range, we find no
 evidence which supports the conclusions of Dragic et al. \cite{Drag}. We estimate that 
there is a 45\% chance of an `interesting' and acceptable DTR pattern following a 
strong FD by statistical fluctuations alone. A Fourier analysis of the DTR pattern 
(`DTR' being, as before, a deviation from a $\pm$10 day running mean) confirms the 
quality of the data. 

The periods for the DTR data detected in the Fourier study are of interest in their own
 right. The periods are consistent with those exhibited by both solar irradiance and 
CR. Other evidence disfavouring CR connections indicate that solar irradiance changes 
are presumably responsible.

Different arguments, too, contribute to the conclusion that CR have a negligible effect
 on the tropospheric climate.

\end{document}